\def\VERSION{4gr}   
\def\DATE{97/Sep/30}

\def\CLFALG{pezz9701}
\def\TITLE{Physical Applications of a Generalized Clifford Calculus}

\def\AUTHOR{William M. Pezzaglia Jr.}
\def\EMAIL{wpezzaglia@scuacc.scu.edu}
\def\HTTP{http://www.well.com/\~billium}

\documentclass{article}		
\usepackage{latexsym}		

\begin{document}


%
\def\HREF#1#2{{#2}}
%
\def\PD#1#2{{\partial {#1} \over \partial {#2}}}
\def\PP#1#2#3{{\partial^2 {#1} \over \partial {#2} \partial {#3}}}
\def\HALF{{1 \over 2}}
\def\EB#1{{\bf e}_{#1}}    
\def\EU#1{{\bf e}^{#1}}     
\def\GB#1{{\bf \gamma}_{#1}}    
\def\GU#1{{\bf \gamma}^{#1}}     
\def\vector#1{\vec{\bf {#1}}}        
\def\NORM#1{\parallel{#1}\parallel}
\def\BAR#1{\overline{\strut{#1}}}
\def\EQN#1{eq.\ (#1)}
\def\HBAR{{\mathchar'26\mkern-9muh}}  
\def\ODOT#1{\stackrel{\circ}{{#1}}}   
\def\DD{{\cal D}}
\def\INNER{\bullet}	

\begin{titlepage}
\def\REFERENCE{Summary of the talk given in the {\it Analysis of Dirac
Operators} section, at the meeting of the {\it International Society for
Analysis, its Applications and Computation},
to appear in {\it the Proceedings of the first annual ISAAC Conference
at the University of Delaware, June 2-6, 1997}.
(http://poseidon.math.udel.edu/isaac/conferen/congr97.html)}

\title{{\bf \TITLE\thanks{REFERENCE}} \\
(Papapetrou equations and metamorphic curvature)} 

\author{{\Large\bf \AUTHOR}
\thanks{Email: \EMAIL}
\\ Department of Physics \\ Santa Clara University
\\ Santa Clara, CA 95053}

\maketitle
\thispagestyle{empty}

\begin{abstract}
A generalized Clifford manifold is proposed in which there are coordinates
not only for the basis vector generators, but for each element of the
Clifford group, including the identity scalar.  These new quantities are
physically interpreted to represent internal structure of matter
(e.g. classical or quantum spin).  The generalized Dirac operator
must now include differentiation with respect
to these higher order geometric coordinates.  In a Riemann space, where the
magnitude and rank of geometric objects are preserved under displacement,
these new terms modify the geodesics.  One possible physical interpretation
is natural coupling of the classical spin to linear motion, providing a new
derivation of the Papapetrou equations.  A generalized curvature is proposed
for the Clifford manifold in which the connection does not preserve the rank
of a multivector under parallel transport, e.g. a vector may be ``rotated''
into a scalar.



\end{abstract}

\end{titlepage}
\pagestyle{myheadings}

\section*{I. Introduction}
To break out of the limitations imposed by habitual thinking, we propose a
different mathematical model for the structure of spacetime.  We adopt the
concept proposed by Chisholm\cite{Chisholm11} of a curved spacetime with 
a Clifford structure imposed upon it (in order to describe quantum particles
with spin in a curved space).  However we take the concept much further in
proposing that it applies to classical physics (non-quantum) and that there
are coordinates for each multivector element.  In section III we
consider derivatives with respect to bivector geometry which leads to 
a new form of the geodesic which is useful in describing the spinning
particle problem.  A new type of curvature is proposed in section IV
in which the connection can modify a vector into a bivector, yielding
a new law of motion for the spinning particle in a gravity field.

\section*{II. The Clifford Manifold}
We propose that space is really a {\it pandimensional continuum}.  In other
words, it is made up of points, lines, planes, volumes etc., all together.
Instead of a vector manifold, we have a ``Clifford Manifold'',
or if you like a {\it polyvector manifold}.  True, there may only be
4 independent tangent basis vectors, but there will also
be 6 tangent basis planes, a tangent scalar, pseudoscalar, and 4
tangent trivectors (pseudovectors), a total of 16 elements.
For our discussion we will not use the full 16
degrees of freedom, but use simpler examples that only utilize the vector
and bivector features.

\subsection*{A. Review of Clifford Algebra}
We recommend the standard references on the subject for complete coverage.
Here we present only brief definitions for notational clarity and some 
specialized equations.
In a non-orthonormal curved space, the basis vectors
(generators of the Clifford algebra) obey,
$$\GB{\mu}\INNER\GB{\nu}\equiv \HALF
\{\GB{\mu},\GB{\nu}\} \equiv 
\HALF \left( \GB{\mu}\GB{\nu}+\GB{\mu}\GB{\nu} \right)
= g_{\mu\nu}, \eqno(1a)$$
where $g_{\mu\nu}$ is {\it metric tensor} which may be a function
of position.  The direct or {\it Clifford product}
of two different basis vectors yields a new object, called a
(basis) bivector.  This is the basis element (2-vector)
for the plane spanned by the
two basis vectors which we can notate,
$$\GB{\mu\nu}=-\GB{\nu\mu} \equiv \GB{\mu}\wedge \GB{\nu}\equiv
\HALF [\GB{\mu},\GB{\nu}]\equiv \HALF
\left( \GB{\mu}\GB{\nu}-\GB{\mu}\GB{\nu} \right). \eqno(1b)$$
The product of three basis vectors is called a trivector.  Assuming we are
in a four dimensional space, then there is only one quadvector, produced
by the product of all four elements.

The unique property of Clifford algebra is the
ability to add together different rank geometries (e.g. scalar plus 
bivector).  This creates a generalized type of entity which is often
called a `multivector' in the literature.  However this name which
literally means `multiple vector'
is ambiguous as it could simply mean an n-vector where $n>1$.
We propose instead the term {\it polyvector} to mean that it
is a combination of a p-vector with a q-vector (where $p\ne q$).
The formulation of standard physics of course does not allow for the
addition of a p-form to a q-form.
We propose to label the formulation of physical theories which
exploit the use of polyvectors as {\it polydimensionalism}, literally the
use of many different dimensional geometric elements in one equation.
As an example, consider the motion of a charged particle in
an electromagnetic field.  We define the {\it momentum polyvector} as the 
vector momentum plus bivector spin angular momentum,
$${\cal M}\equiv {1 \over \lambda} p^\mu \GB{\mu} 
+ {1 \over 2\lambda^2} S^{\mu\nu}\GB{\mu\nu}, \eqno(2a)$$
$$p^\mu = m {dx^\mu \over d\tau}, \eqno(2b) $$
$$p_\mu S^{\mu\nu} = 0. \eqno(2c)$$
The Weysenhoff condition \EQN{2c} insures the spin to be
a simple bivector, which is purely spacelike in the rest frame of the
particle.  The scale factor $\lambda$ is a universal length constant that
will be defined in the next section, and the proper time $d\tau$ is
given below in \EQN{7}.  The Einstein summation convention on repeated
indices will be used throughout the paper.
With charge-to-mass ratio ${e/m}$, the polydimensional equation 
of motion is simply,
$$\dot{\cal M}\equiv
{d {\cal M} \over d\tau} = {e \over 2m} [{\cal M},{\bf F}], \eqno(3a)$$
where ${\bf F}=\HALF F^{\mu\nu}\GB{\mu\nu}$ is the electromagnetic field
tensor.  The vector and bivector parts of \EQN{3a}
describe the linear motion and spin motion respectively,
$$\dot{p}^\beta \equiv {d p^\beta \over d\tau}
=\left( {e \over m} \right) p_\alpha F^{\alpha\beta}, \eqno(3b)$$
$$\dot{S}^{\mu\beta}\equiv {d S^{\mu\beta} \over d\tau}
=\left( {e \over2 m} \right)
\left( F^\mu_{\ \nu}S^{\nu\beta}-F^\beta_{\ \nu}S^{\nu\mu}
\right). \eqno(3c)$$
This is a somewhat trivial example because there is no coupling between
the spin and momentum.
In tensor form \EQN{3b} and \EQN{3c} do not look really look the same,
while \EQN{3a} shows a possible underlying symmetry between the two
equations.  Is this just notational economy or is there a broader
relationship being revealed which is obscured in tensor form?

\subsection*{B.  Dimensional Democracy}
In quaternionic
analysis, one gives coordinates $x^\mu$ to all 4 elements,
${\bf H}=x^1 {\bf i}+ x^2 {\bf j} + x^3 {\bf k} + x^4 {\bf 1}$.  However, 
when viewed as a Clifford algebra, the three imaginaries are NOT all the
same `rank' of geometry.  Two of them are basis vectors (1-vectors),
and the third is really a basis plane (2-vector), while the fourth
element is a scalar (0-vector).  Hence we
have given coordinates to ALL the geometric elements.

The standard formulation of physical theories might be called a
{\it vector oligarchy} as coordinates are only associated
with the basis vectors (1-vectors) of the spacetime manifold.
Einstein's relativity for example represents the structure of
space-time with a four-dimensional vector manifold.  The coordinates of an
{\it event} $x^\mu({\cal P})$ for $\mu =1,2,3,4$.
The fourth dimension: $x^4=ct$ is proportional to `time' $t$, where the
speed of light $c$ is the universal constant which converts time into
equivalent spatial units.  From the differential of the event point:
$d{\cal P}\equiv \EB{\mu} dx^\mu$ the tangent basis vectors
can defined by applying the chain rule of differentiation,
$$\EB{\mu}\equiv \PD{\cal P}{x^\mu}. \eqno(4)$$
In our practical world, the quantity $x^\mu$ (more properly a finite
interval $\Delta x^\mu$)
is quantified in terms of units, such as meters or feet.  However the
quantity $d{\cal P}$ is purely a geometric interval which has no units.
Thus by \EQN{4} the basis vectors must have units of inverse length!
If we change from feet to inches these basis vectors will
need to be decreased by a corresponding factor of twelve.  The relationship
of these tangent vectors to the purely algebraic basis set 
$\GB{\mu}$ (which has no units), is hence:
$\GB{\mu}\equiv \EB{\mu} \lambda$, 
where $\lambda$ is the scale constant associated with the particular unit
system in which we measure $x^\mu$.

We will define the Clifford manifold in terms of these four tangent basis
vectors.  Let us define the set of 16 elements:
$\{{\bf E}_A\}=\{1,\EB{\mu},\EB{\mu\nu},\EB{\alpha\mu\nu},
\EB{\alpha\beta\mu\nu} \}$
[the notation implies antisymmetry on the indices].  We
now propose that each basis element ${\bf E}_A$
has a coordinate $q^A$ associated with it.  The bivector coordinate:
$a^{\mu\nu}=-a^{\nu\mu}$ would for example have the units of area.
An event ${\Sigma}$ in pan-dimensional space has a unique set of
coordinates $q^A ({\Sigma})$.  The {\it pandimensional differential}
is hence,
$$
d{\Sigma}  \equiv {\bf E}_A dq^A $$
$$={\bf 1}d\kappa + \EB{\mu}dx^\mu + \HALF \EB{\mu\nu}
da^{\mu\nu}+{1 \over 3!} \EB{\mu\nu\omega}
\epsilon^{\mu\nu\omega\alpha}dy_\alpha +
{1 \over 4!} \EB{\mu\nu\omega\alpha}
\epsilon^{\mu\nu\omega\alpha} d\sigma, \eqno(5)$$
where $\epsilon^{\mu\nu\omega\alpha}$ is the totally antisymmetric
tensor.
By construction, \EQN{5} has no units, it is a pure geometric
infinitessimal interval.

For the remainder of the paper we will restrict our treatment to only
the vector and bivector coordinates.  A finite 
interval of the vector coordinates implies a displacement from event
${\Sigma}$ to ${\Sigma}^\prime$.  The simultaneous bivector displacement
might represent the amount by which the path of the particle deviated
from the straight line path, or perhaps the area traced
out by its spinning motion.  The momentum of a particle defined in
\EQN{2b} is the rest mass $m$ times the rate of 
displacement with respect to the affine parameter, called the 
{\it proper time} in relativity, defined below in \EQN{7}.  A possible
interpretation of 
the bivector coordinate would be the quantity for which the
spin angular momentum of the particle is given in analogy to \EQN{2b},
$$S^{\mu\nu}\equiv m {d a^{\mu\nu} \over d\tau}. \eqno(6)$$
The momentum polyvector \EQN{2a} can now be expressed compactly as:
${\cal M} \equiv m {d {\Sigma} \over d\tau}$.

\subsection*{C.  Quadratic Forms}
The inner product on a vector space induces the quadratic form:
$\NORM{d{\cal P}}^2 \equiv dx^\mu dx^\nu \ {g_{\mu\nu} / \lambda^2}$
where we recall the metric tensor is:
$g_{\mu\nu}\equiv \GB{\mu} \INNER \GB{\nu} =
\lambda^2\ \EB{\mu}\INNER \EB{\nu}$.
There is an affine parameter associated with the quadratic form, known as
the {\it proper time} in mechanics (as it is the time that the particle
experiences in its own frame of reference),
$$d\tau \equiv {1 \over c}
\sqrt{s g_{\mu\nu}(x^\alpha)\ dx^\mu dx^\nu}, \eqno(7)$$
which has units of time.  For the $(---+)$ metric the {\it metric
signature} $s$ is defined to be
$s=+1$, while  $s=-1$ for the $(+++-)$ metric.
Although relativity cannot distinguish between these
two cases, the Clifford group associated with each case is inequivalent
which may have consequences in physical theories\cite{Pezz9502}.  Most of
the results in this paper require $s=+1$.
The basic precept of classical mechanics is that out of all the 
possible trajectories connecting two events, nature will ``choose'' the
special path which extremizes the quadratic form of \EQN{7} integrated
over the path.  The solution is a {\it geodesic} in a curved space
(or an ``autoparallel'' if there is torsion).

In classical mechanics, the modulus of the momentum vector,
$${\bf p}^2 = \left( p^\mu \GB{\mu}\right)^2=p^\mu p^\nu g_{\mu\nu}
=p^\mu p_\mu = s(mc)^2, \eqno(8)$$
must be invariant under a change of coordinate system.  The invariant
$m$ is the rest mass of the particle.  Corben\cite{Corben61} and others
have argued that for a spinning particle \EQN{8} gives the 
{\it effective mass}, which is \underline{not} the same as the 
{\it intrinsic rest mass} $m_o$. Note if we naively square the
momentum polyvector efined in \EQN{2a} (in the $s=+1$ signature),
we get a scalar plus trivector,
$${\cal M}^2 = {c^2  \over \lambda^2}I_1 + 
\lambda^{-3}{\bf p}\wedge {\bf S}, \eqno(9a)$$
$$I_1 \equiv {p^\mu p_\mu \over c^2} - {S^{\mu\nu}S_{\mu\nu}
\over 2\lambda^2 c^2}=s m^2 - {S^2 \over \lambda^2 c^2}
\equiv m_o^2, \eqno(9b)$$
$$S^2 \equiv \HALF S^{\mu\nu}S^{\alpha\beta}\ g_{\mu\alpha}
\ g_{\nu\beta}=\HALF S^{\mu\nu}S_{\mu\nu}, \eqno(9c)$$
where \EQN{2c} gives us $S^2\ge 0$.  The scalar part of the modulus of
the momentum polyvector is proportional to
the invariant $I_1$ which Dixon\cite{Dixon70} associates with the
(square of the) intrinsic or ``bare'' rest mass $m_o$.  In other words,
the spin of the particle increases the effective mass $m$ which appears
in equations (2b), (6) and (8).  We might also note that the
modulus of the trivector part of \EQN{9a} yields the second invariant
$I_2$ of Dixon\cite{Dixon70}.  He also interprets the scale constant
$\lambda$ as being possibly the radius of the universe, although we
think it should be the {\it radius of gyration} of a macroscopic body,
or within a geometric factor of the Compton wavelength for an
elementary particle 
(for which the spin would be a fundamental constant).

This suggests a way to generalize \EQN{7} for polydimensional
coordinates.  Keeping only 
the vector and bivector ccordinates of \EQN{5} and
squaring it (in the $s=+1$ signature),
$$\left( d\Sigma \right)^2 \equiv \left({c \over \lambda } \right)^2
d\kappa^2 + (trivector),\eqno(10a)$$
$$d\kappa^2 \equiv d\tau^2 - {da^{\mu\nu}da_{\mu\nu} \over
2c^2 \lambda^2 }. \eqno(10b)$$
The new affine parameter $d\kappa$  is \underline{not} the proper
time $d\tau$ of special relativity, as there is an additional
contribution from the bivector displacement.  We can be consistent with
\EQN{9b}, showing the spin gives an increase in the effective mass,
if we rewrite the definitions
for the momentum and spin in the following way,
$$p^\mu \equiv m_o \ODOT{x}{}^\mu \equiv
m_o{dx^\mu \over d\kappa}=m \dot{x}^\mu , \eqno(11a)$$
$$S^{\mu\nu} \equiv m_o \ODOT{a}{}^{\mu\nu} \equiv
m_o {da^{\mu\nu} \over d\kappa}=
m \dot{a}^{\mu\nu} , \eqno(11b)$$
$$m\equiv m_o {d\tau \over d\kappa} \equiv m_o  
\sqrt{1 +\left({S \over \lambda c m_o}\right)^2}. \eqno(11c)$$

\section*{III.  Generalized Clifford Analysis}
With the introduction of the bivector coordinate, we are ontologically 
committed to consider the meaning of a derivative with respect to this
coordinate.

\subsection*{A. Differential Multiforms}
The standard differential operator ${\bf d}$ applied to the vector
coordinate gives ${\bf d}x^\mu = dx^\mu$ as desired, where $dx^\mu$
denotes an infinitessimal quantity.  Note however that the {\it vector
differential}: $d{\bf x}\equiv \EB{\mu}dx^\mu$ must be defined as it can
NOT be constructed by an application of
${\bf d}$ on ${\bf x}=x^\mu \EB{\mu}$.  Similarly, the {\it bivector
differential}:
$d{\bf a}\equiv \HALF \EB{\mu}\wedge \EB{\nu}da^{\mu\nu}$ must be 
defined rather than constructed.  When integrated, this entity should
yield the area displacement.  Hence $d{\bf a}$ is really a
{\it Leibniz form of second order}, the form remaining after the
magnitude of area has gone to zero.  Hence we wish to associate
$da^{12}=dx dy$, which can not be constructed by applying the standard
${\bf d}$ on $a^{\mu\nu}$.  We have instead,
$${\cal D}a^{\mu\nu}=da^{\mu\nu}=-da^{\nu\mu}
\equiv dx^\mu dx^\nu,\ \ (\mu<\nu), \eqno(12a)$$
where the operator ${\cal D}$ is defined below in \EQN{12b}.

It remains to provide a geometric interpretation to the pandimensional
differential of \EQN{5}.  Let us consider the restricted case of the
polydimensional differential which only has the vector and bivector
portion.  We might interpret the finite polydimensional displacement
as follows.  The vector part tells the straight line vector displacement
connecting the endpoints.  The bivector part describes the amount by which
the actual path deviated from the straight line (expressed in terms of the
area enclosed between the actual path and the straight path).

The physical application for which the polydimensional differential may
be useful is in describing the spinning particle.  Corben\cite{Corben61} for
example states that the elementary particle travels in a helical path
about the motion of its center of mass, the helical motion is (in part)
seen as the `spin' of the particle.  We could hence let the vector part
of the differential describe the center of mass motion, and the bivector
part the area swept out by the helical motion.

\subsection*{B.  The Clifford Polydifferential Operator}
Consider the function $f(x^\mu, a^{\mu\nu})$ which has a value at
point $x^\mu$ that will vary according to the path ${\cal P}$ taken
from the origin, which in turn is described by the bivector coordinate.
We introduce the {\it Clifford Polydifferential Operator} which includes
the higher order coordinates (for our treatment we will consider only
the vector and bivector coordinates),
$${\bf \cal D}f \equiv dq^A \PD{f}{q^A}=dx^\mu \PD{f}{x^\mu}
+\HALF da^{\mu\nu}\PD{f}{a^{\mu\nu}}. \eqno(12b)$$

The problem is to provide interpretation to the derivative with respect
to the bivector coordinate.  The total change in
the function from the origin to the endpoint along the path ${\cal P}$
is the integral of above.  This can be partitioned into the integral
along the straight (geodesic) path ${\cal P}_o$ plus the closed integral
which is the true path ${\cal P}$ minus the return by path ${\cal P}_o$,
$$\triangle f \equiv \int {\bf \cal D}f = \Delta {\bf x}\INNER \nabla f
+\oint {\bf d}f, \eqno(12c)$$ 
Inserting \EQN{12b} into the left integral, matching terms and applying
Stokes theorem for differential forms we get,
$$\HALF \int da^{\mu\nu} \PD{f}{a^{\mu\nu}} =\oint {\bf d}f
=\HALF \int da^{\mu\nu} [\partial_\nu,\partial_\mu ]f. \eqno(12d)$$
We therefore propose the association of the bivector derivative with,
$$\PD{f}{a^{\mu\nu}}\equiv [\partial_\nu,\partial_\mu ]f=
\left[ \PD{}{x^\nu},\PD{}{x^\mu} \right]f. \eqno(12e)$$

This gives a new non-standard result for the exact differential of a
basis vector in curved space,
$${\bf \cal D}\EB{\mu}=\left( dx^\alpha \ \Gamma_{\alpha\mu}{}^\nu
-\HALF da^{\alpha\beta}\ R_{\alpha\beta\mu}{}^\nu
\right) \EB{\nu}, \eqno(13a)$$
$$\Gamma_{\alpha\mu}{}^\nu \equiv {\bf e}^\alpha \INNER
\partial_\mu \EB{\nu}, \eqno(13b)$$
$$R_{\alpha\beta\mu}{}^\nu \equiv {\bf e}^\nu \INNER
\left[ \partial_\alpha, \partial_\beta \right]\EB{\mu}=
-{\bf e}^\nu \INNER \PD{\EB{\mu}}{a^{\alpha\beta}}. \eqno(13c)$$
Not only does \EQN{13a} contain the usual affine connection 
$\Gamma_{\alpha\mu}^{\ \nu}$, it also contains the Riemann curvature
tensor $R_{\alpha\beta\mu}{}^\nu$. 
The connection of a bivector can be derived from \EQN{13a},
$${\cal D} \EB{\mu\nu} = \left(dx^\alpha \ \Gamma_{\alpha\omega}{}^\sigma
-\HALF da^{\alpha\beta}\ R_{\alpha\beta\omega}{}^\sigma \right)
\delta^{\omega\kappa}_{\mu\nu}\ \EB{\sigma\kappa}. \eqno(13d)$$

\subsection*{C. Spinning Particles in Curved Space}
Galileo's famous experiment asserted that big balls fall at the same rate
as small balls.  The more general statement is known as the {\tt EEP}
(Einstein equivalence principle), that  all particles follow the same
geodesic path, independent of the mass or internal structure.
Specifically, a spinning mass
should fall at the same rate as a non-spinning one.  However, in a 
landmark paper, Papapetrou\cite{Papa} showed that spinning masses will
deviate from geodesics.  His derivation was based upon taking a 
macroscopic mass, expanding in moments about the center, and looking at
the geodesic deviations.  The basic result is that the deviation is
proportional to the spin of the particle, the curvature of the space and
inversely proportional to the mass.  It is presumed that this result is
valid for fundamental particles, but there is no classical theory that
yields the result.
The difficulty is that there has not been a satisfactory theory which 
derives these equations from a simple ``least action'' principle based 
upon purely geometric concepts such that Einstein used.  There
have been far too numerous attempts, a good recent review is given by
Frydryszak\cite{Frydryszak}.

By introducing the bivector coordinates, we believe that we can now
present a new derivation of the Papapetrou equations.  They are the
{\it polygeodesic} paths, which extremize the polydimensional quadratic
form of \EQN{10b}.  The proof of this statement is too long to show
here, hence will be presented in a future paper\cite{Pezz9702}.
The main difficulty is
that in applying Lagrange's equations, one finds that the variation of
the path does not commute with the derivative with respect to the 
new affine parameter,
$${d \delta x^\mu \over d\kappa}\ne \delta \left({dx^\mu \over d\kappa}
\right). \eqno(14a)$$

We can however show here that the resulting equations of motion can be
expressed in the very compact form  of a {\it polygeodesic}.  This
is defined as the the polycurve
$\Sigma(\kappa)$ that ``parallel transports'' its tangent polyvector
along itself,
$$0=\ODOT{{\cal M}}\equiv{d {\cal M} \over d\kappa}=
\PD{\cal M}{\kappa}+\ODOT{x}{}^\mu\ \PD{\cal M}{x^\mu}
+\HALF \ODOT{a}{}^{\mu\nu}\ \PD{\cal M}{a^{\mu\nu}}. \eqno(14b)$$
Explicitly substituting \EQN{13ad} as needed, one can split out 
the following equations for the momentum and spin respectively,
$$ 0=\ODOT{\bf p}\equiv
{d \left( p^\mu \EB{\mu}\right) \over d\kappa}=\left(\ODOT{p}{}^\mu
+p^\nu \left[ \ODOT{x}{}^\beta\ \Gamma_{\beta\nu}{}^{\mu} 
+\HALF  \ODOT{a}{}^{\alpha\beta}
\ R_{\alpha\beta\nu}{}^{\mu}\right] \right)\EB{\mu}, \eqno(14c)$$
$$0=\ODOT{\bf S}\equiv \HALF
{d \left( S^{\alpha\beta} \EB{\alpha\beta}\right) \over d\kappa}$$
$$=\Bigl\{\ODOT{S}{}^{\alpha\beta} + \ODOT{x}{}^\nu
\left(S^{\sigma\beta}\Gamma_{\nu\sigma}^{\ \alpha}
+S^{\alpha\sigma}\Gamma_{\nu\sigma}^{\ \beta} \right)
+\HALF \ODOT{a}{}^{\mu\nu} \left( S^{\sigma\beta} 
R_{\mu\nu\sigma}^{\ \ \ \alpha}+S^{\alpha\sigma}
R_{\mu\nu\sigma}^{\ \ \ \beta} \right) 
\Bigr\}{\EB{\alpha\beta} \over 2}, \eqno(14d)$$
These are equivalent to the Papapetrou equations\cite{Papa} except that the
`open dot' refers to differentiation by the new affine parameter $d\kappa$
of \EQN{10b} instead of the old proper time $d\tau$ of \EQN{7}.

\section*{IV.  Geo-Metamorphic Curvature}
We propose a radical concept of curved space in which
a vector may be `bent' into a bivector.  These ideas were first 
introduced in an earlier paper\cite{Pezz9601}.  We might alternatively
call this polymetamorphic curvature, or more generally {\it pan-dimensional
curvature}.

\subsection*{A.  Automorphism Invariance}
A principle of special relativity is that the laws of physics must be
of the same form in all frames of reference that differ only by a
(global) transformation generated from the Lorentz group.  Note that
scalars such as \EQN{7}, \EQN{8}, \EQN{9b} and even \EQN{10b}
are invariant under these transformations.  Lets consider however the
broader class of algebra automorphisms which just leave \EQN{9b} invariant.
They would be of the type that interchange some of the basis vectors
with some of the basis bivectors, but in such a way that the overall
algebra is preserved.  The polymomentum vector would hence transform:
${\cal M}^\prime = {\cal Q} {\cal M} {\cal Q}^{-1}$ where,
$${\cal Q}=\exp\left(\HALF \GB{\mu}\psi^\mu + 
{1 \over 4} \GB{\mu\nu}\phi^{\mu\nu}\right).  \eqno(15)$$
The parameters
$\phi^{\mu\nu}$ represent the Lorentz transformations, which 
preserve the rank of the geometry, while $\psi^\mu$ generates the
transformations that will exchange vectors with bivectors, and
the pseudoscalar with one of the trivectors.

This leaves \EQN{9b} invariant, and most important leaves
\EQN{14b} invariant in form. We
have hence proposed\cite{Pezz9601} a 
generalization of the covariance principle, that
the laws of physics must be the same form in frames of reference that
differ only by a global (restricted) automorphism transformation.
This also means that there is no absolute `direction' to which one
can assign the geometry of `vector'.  What is a vector in one frame
of reference may be a bivector in another.  Taken a step further, 
observers in the different frames will disagree on the interpretation
of the phenomena.  They will both agree on the bare mass given by 
the invariant of \EQN{9b}, but will disagree on the amount of the particle's
spin and momentum.  This is because the coordinate transformation between
the two frames is such that the vector coordinate $z^\alpha$
in one frame may be a function
of both the vector AND bivector coordinate of the other frame:
$z^\alpha (x^\beta, a^{\mu\nu})$.  For example, the vector displacement in
a rotating frame is the vector displacement of the non-rotating frame minus
the contribution of the rotation (described by a bivector).  The careful
reader might note that \EQN{3a} is NOT invariant under our automorphism
transformation, but will acquire an additional term which possibly could
be interpreted as giving the photon mass.

\subsection*{B.  Polymorphically Connected Space}
In relativistic quantum mechanics, Crawford\cite{Crawford} has proposed
a {\it local automorphism invariant principle} as a method to generate
a gauge theory of gravity and possible unified field theory.  The basic
feature he introduced was a {\it Clifford connection} on the abstract
Dirac matrix (Clifford) algebra.  The analogous form on our very concrete 
geometric Clifford algebra would be something like,
$${\cal D}\EB{\mu}=dx^\alpha \left(\Gamma_{\alpha\mu}{}^\nu \ \EB{\nu}
+\HALF \Xi_{\alpha\mu}{}^{\nu\sigma}\ \EB{\nu\sigma} \right). \eqno(16)$$
We might call $\Xi_{\alpha\mu}{}^{\nu\sigma}$
a {\it metamorphic connection} because under parallel transport
it can change a vector into a bivector!  This is a logically incomplete
theory however, because only the basis vectors have been given
a coordinate.  Hence the metamorphic connection cannot be expressed
in terms of a coordinate transformation.

We introduce a {\it polymorphic coordinate transformation} from
a flat cartesian frame with coordinates $(z^j, \phi^{jk})$ to
polymorphically curved coordinates $(x^\alpha, a^{\mu\nu})$
via an {\it anholonomic} local transformation,
$$dz^j \equiv h^j_\alpha dx^\alpha + 
\HALF K^j_{\mu\nu}da^{\mu\nu}, \eqno(17a)$$
$$d\phi^{jk}\equiv C^{jk}_\alpha dx^\alpha 
+ \HALF H^{jk}_{\mu\nu}da^{\mu\nu}. \eqno(17b)$$
The coefficient $h^j_\alpha$ is the familiar vierbein (tetrad).  The 
entire set of coefficients $\{h^j_\alpha , K^j_{\mu\nu},
C^{jk}_\alpha,H^{jk}_{\mu\nu} \}$ , has been
collectively called the {\it geobeins} or
``geometry-legs''\cite{PezzMex}.

The generalized polymorphic Clifford connection is hence,
$${\cal D}\EB{\mu}=dx^\alpha \left(\Gamma_{\alpha\mu}{}^{\nu}\ \EB{\nu}
+\HALF \Xi_{\alpha\mu}{}^{\nu\sigma}\ \EB{\nu\sigma} \right)
-\HALF da^{\alpha\beta}\left(R_{\alpha\beta\mu}{}^{\nu}\ \EB{\nu}
+\HALF Q_{\alpha\beta\mu}{}^{\nu\sigma}\ \EB{\nu\sigma} \right),
\eqno(18a)$$
where the coefficients can now be written in terms of the geobeins.
For example,
$$\Gamma_{\alpha\mu}^{\ \nu}=h^\nu_{\ j}h_{\mu,\alpha}^{j} +\HALF
K^\nu_{\ ij}C_{\mu,\alpha}^{ij}
=\PD{x^\nu}{z^j}\PP{z^j}{x^\alpha}{x^\mu}
+\HALF \PD{x^\nu}{\phi^{ij}}\PP{\phi^{ij}}{x^\alpha}{x^\mu}, \eqno(18b)$$
$$\Xi_{\alpha\mu}^{\ \ \nu\sigma}=C_j^{\nu\sigma}h^j_{\mu,\alpha}
+\HALF H_{ij}^{\nu\sigma} C_{\mu,\alpha}^{ij}
=\PD{a^{\nu\sigma}}{z^j}\PP{z^j}{x^\alpha}{x^\mu}
+\HALF \PD{a^{\mu\sigma}}{\phi^{ij}}
\PP{\phi^{ij}}{x^\alpha}{x^\mu}, \eqno(18c)$$
where the first term in \EQN{18b} is the standard part.
In deriving the connection
for the bivector, one discovers that the Leibniz rule does not hold
for the Clifford polydifferential operator $\DD$ over the
wedge or dot product, e.g. $\DD\left(\EB{\mu}\wedge\EB{\nu}\right) \ne
\left(\DD\EB{\mu}\right)\wedge\EB{\nu}
+\EB{\mu}\wedge\left(\DD\EB{\nu}\right)$.  However it
works fine for the Clifford (direct) product.  Hence,
$$\DD\left(\EB{\mu}\wedge\EB{\nu}\right)=
\HALF \left[\left(\DD\EB{\mu}\right),\EB{\nu}\right]
+\HALF\left[\EB{\mu},\left(\DD\EB{\nu}\right)\right]$$
$$=\delta^{\omega\kappa}_{\mu\nu}
\left[ \left(dx^\alpha \Gamma_{\alpha\omega}{}^\sigma
-\HALF da^{\alpha\beta} R_{\alpha\beta\omega}{}^\sigma \right)
 \EB{\sigma\kappa}
+\left( dx^\alpha \Xi_{\alpha\omega}{}^{\xi\sigma}
-\HALF da^{\alpha\beta} Q_{\alpha\beta\omega}{}^{\xi\sigma}\right)
g_{\kappa\sigma}\EB{\xi}\right] , \eqno(19)$$
where again $\Xi_{\alpha\omega}{}^{\xi\sigma}$ and
$Q_{\alpha\beta\omega}{}^{\xi\sigma}$ are the metamorphic connections.

\subsection*{C.  Metamorphic Polygeodesics}
Parallel transporting the momentum vector around a closed loop might
have it return as a pure bivector due to the metamorphic connection!
However, if we transport the momentum polyvector, then it will return
as the same form  but with a reshuffling of the momentum and
spin portions, subject to the constraint that the modulus
of \EQN{9b} be unchanged.
Physically this would appear as additional forces due to the coupling
between the spin and momentum via the metamorphic connection.
Although \EQN{14b} is still valid, \EQN{14cd} will be modified with
terms that couple the equations.  In particular \EQN{14c} becomes,
$$\ODOT{p}{}^\mu +p^\nu \left( \ODOT{x}{}^\beta 
\ \Gamma_{\beta\nu}^{\ \mu} +\HALF  \ODOT{a}{}^{\alpha\beta}
\ R_{\alpha\beta\nu}{}^\mu\right)
+\HALF S^\omega_{\ \sigma}\left(\ODOT{x}{}^\alpha 
\ \Xi_{\alpha\omega}^{\ \ \mu\sigma} + \HALF\ODOT{a}{}^{\alpha\beta}
\ Q_{\alpha\beta\omega}{}^{\mu\sigma}\right)=0 , \eqno(20a)$$
Note that the metamorphic connection $\Xi_{\alpha\omega}^{\ \ \mu\sigma}$
enhances the Papapetrou term, while the other metamorphic 
connection $Q_{\alpha\beta\omega}{}^{\mu\sigma}$
adds a pure second order spin interaction.  The generalization of
\EQN{14d} can be generated in the same way, but will be left to a
more detailed future paper\cite{Pezz9702}.

\section*{V.  Summary}
In introducing {\it Dimensional Democracy} we have given the bivector a
coordinate and shown its utility in treating the spinning particle
problem.  In adopting {\it automorphism invariance} we find that the
concept of what is a vector depends upon the observer's frame of 
reference, leading to demanding that physical laws are covariant
under this transformation.  This allows us to derive
Crawford's {\it Clifford Connection}\cite{Crawford} in a classical
context as a local polymorphic coordinate transformation.  Most 
important the themes we have introduced provide an entirely new 
general way in which to apply Clifford calculus to physical theories.

\end{document}